\documentclass[a4paper,12pt]{article}
\usepackage{amssymb}
\usepackage{amsfonts}
\usepackage{amsmath}
\usepackage{mathrsfs}
\usepackage{cite}

\renewcommand{\thefootnote}{\fnsymbol{footnote}}

\csname @addtoreset\endcsname{equation}{section}

\newcommand{\eq}{\begin{equation}}
\newcommand{\feq}{\end{equation}}
\newcommand{\eqn}{\begin{eqnarray}}
\newcommand{\feqn}{\end{eqnarray}}
\newcommand{\arr}{\begin{eqnarray*}}
\newcommand{\farr}{\end{eqnarray*}}

\newcommand{\be}[3]{\begin{equation}  \label{#1#2#3}}
\newcommand{\eeq}{\end{equation}}
\newcommand{\ba}{\begin{array}}
\newcommand{\ea}{\end{array}}
\newcommand{\bea}[3]{\begin{eqnarray}  \label{#1#2#3}}
\newcommand{\eea}{\end{eqnarray}}

\font\mybb=msbm10 at 12pt
\def\bb#1{\hbox{\mybb#1}}

\def\bR {\bb{R}}

\begin{document}

\baselineskip=18pt
\parskip=4pt

\begin{titlepage}
\begin{flushright}
AEI-2004-002\\
IFUM-779-FT\\
hep-th/0401239
\end{flushright}

\vspace{.3cm}
\begin{center}
\renewcommand{\thefootnote}{\fnsymbol{footnote}}
{\Large \bf Black holes in  G\"odel-type universes \\[4mm] 
with a cosmological constant}

\vskip 15mm

{\large \bf {Klaus Behrndt$^1$\footnote{behrndt@aei.mpg.de}
and Dietmar Klemm$^2$\footnote{dietmar.klemm@mi.infn.it}}}\\
\renewcommand{\thefootnote}{\arabic{footnote}}
\setcounter{footnote}{0}
\vskip 10mm
{\small
$^1$Max-Planck-Institut f\"ur Gravitationsphysik, Albert-Einstein-Institut,\\
Am M\"uhlenberg 1, 14476 Golm, Germany\\
\vskip 7mm
$^2$Dipartimento di Fisica dell'Universit\`a di Milano\\
and\\
INFN, Sezione di Milano,\\
Via Celoria 16, 20133 Milano, Italy\\
}
\end{center}
\vspace{1cm}
\begin{center}
{\bf Abstract}
\end{center}
{\small We discuss supersymmetric black holes embedded in a
G\"odel-type universe with cosmological constant in five
dimensions. The spacetime is a fibration over a four-dimensional
K\"ahler base manifold, and generically has closed timelike
curves. Asymptotically the space approaches a deformation of AdS$_5$,
which suggests that the appearance of closed timelike curves should
have an interpretation in some deformation of $D=4$, ${\cal N}=4$
super-Yang-Mills theory.\\
Finally, a G\"odel-de~Sitter universe is also presented and its
causal structure is discussed.}

\end{titlepage}


\section{Introduction}

For specific cases we have already a fairly good understanding of
(vacuum) geometries in which string theory can be embedded without
breaking supersymmetry. A general picture is however still missing,
but one step in this direction is the classification of supersymmetric
bosonic field configurations of lower-dimensional supergravity,
obtained by compactification of string theory (or M-theory). For the
minimal supergravity in five dimensions this was recently done in
\cite{Gauntlett:2002nw,Gauntlett:2003fk}\footnote{For a systematic
classification of BPS supergravity solutions in other dimensions
cf.~\cite{Tod:pm}.}, where the BPS solutions were classified by a
Killing vector field, which is always present due to supersymmetry. In
fact, unbroken supersymmetry requires the existence of at least one
Killing spinor, which in turn implies the existence of a Killing
vector. This Killing vector is constructed as fermionic bi-linear, and
can be null or timelike, but not spacelike. The null case describes
pp-wave-type solutions, whereas examples with a timelike Killing
vector are the BPS black holes \cite{Gibbons:xt}.

There is another class of BPS solutions with a timelike Killing
vector, that are however neither asymptotically flat nor
anti-de~Sitter as in the case of gauged supergravity. These are the
G\"odel-type solutions of \cite{Gauntlett:2002nw}, which are
pathological in the sense that they exhibit closed timelike curves
(CTCs), which are not shielded by any horizon. Many attempts have been
made to understand or to cut-off the regions with CTCs by holographic
screens or by appropriate probes
\cite{Boyda:2002ba,Drukker:2003sc,Hikida:2003yd,Brecher:2003rv,Israel:2003cx,
Brace:2003rk}, but so far a deeper
understanding of this phenomenon is still missing. One interesting
observation is the link of the ungauged case to an integrable model
and the appearance of a pole in the partition function indicating a
phase transition \cite{Russo:1994cv,Brace:2003st}.  In gauged
supergravity on the other hand, all CTCs disappear if the cosmological
constant is sufficiently large \cite{Behrndt:2003gc}. The situation
here is reminiscent of the rotating BMPV black hole \cite{Breckenridge:1996is}
in asymptotically flat spacetime, where CTCs are present only in the
over-rotating case, but disappear if the mass (for fixed angular momentum)
becomes large enough \cite{Gauntlett:1998fz,Gibbons:1999uv}.

In this paper, we want to discuss in more detail the interplay between
(rotating) black holes and G\"odel solutions. We are particularly
interested in black holes embedded in G\"odel universes with
cosmological constant, that approach asymptotically a deformation of
AdS$_5$. The AdS/CFT correspondence opens then the possibility to
relate the appearance of closed timelike curves in the bulk to
properties (like loss of unitarity) of a dual field theory residing on
the deformed boundary of the five-dimensional spacetime.

In detail, the remainder of this paper is organized as follows: In
section \ref{susygoedelBH}, we construct various BPS black hole
solutions embedded in a G\"odel-AdS spacetime. We discuss their causal
structure and point out a possible holographic interpretation of the
appearance of closed timelike curves in the AdS/CFT correspondence.
In section \ref{moregen} a more general class of BPS solutions is
constructed. These solutions are given in terms of a fibration over a
four-dimensional K\"ahler base manifold which is a complex line bundle
over a two-dimensional surface of constant curvature. They include a
G\"odel-type deformation of the rotating AdS black holes obtained
recently in \cite{Gutowski:2004ez} as a special subcase. Finally, in
section \ref{GdS} we present a G\"odel-de~Sitter universe and analyze
some of its physical properties.

\section{Supersymmetric G\"odel-AdS black holes}

\label{susygoedelBH}

\subsection{Construction of the supergravity fields}

Gauntlett and Gutowski \cite{Gauntlett:2003fk} classified all
supersymmetric solutions of minimal gauged supergravity in five
dimensions, with bosonic action
\begin{equation}
S = \frac{1}{4\pi G}\int \left(-\frac 14[^5R - 2\Lambda]\ast{\mathrm 1}
    -\frac 12 F\wedge\ast F - \frac{2}{3\sqrt 3}F\wedge F\wedge A \right)\,.
    \label{action}
\end{equation}
The solutions fall into two classes, depending on whether the Killing vector
constructed from the Killing spinor is timelike or null. Let us consider the
former class. In order to make our paper self-contained, we briefly review
the results of \cite{Gauntlett:2003fk} for the timelike case.
The line element can be written as
\begin{equation}
ds^2 = f^2(dt + \omega)^2 - f^{-1}h_{mn}dx^m dx^n\,,
\end{equation}
where $h_{mn}$ denotes the metric on a four-dimensional K\"ahler base
manifold ${\cal B}$ and
\begin{equation}
f = -\frac{2\chi^2}{R}\,,
\end{equation}
where $R$ is the scalar curvature of ${\cal B}$ and $\chi$ is related
to the cosmological constant by $\chi^2 = 2\Lambda$\footnote{With
mostly minus signature, positive $\Lambda$ corresponds to AdS and
negative $\Lambda$ to dS.}.  The one-form $\omega$ is determined by
\begin{equation}
f d\omega = G^+ + G^- \label{omega}
\end{equation}
where $G^+$ is a self-dual two-form on the base manifold,
given by
\begin{equation}
G^+_{mn} = -\frac{\sqrt 3}{\chi}({\cal R}_{mn} - \frac 14 R X^{(1)}_{mn})\,,
\end{equation}
with ${\cal R}$ the Ricci form and $X^{(1)}$ the K\"ahler form of
${\cal B}$.  On the other hand $G^-$ is an anti-self-dual two-form and
is decomposed as
\begin{equation}
G^- = \lambda^1 X^{(1)} + \lambda^2 X^{(2)} + \lambda^3 X^{(3)}\,,
\end{equation}
where $X^{(2)}$ and $X^{(3)}$ are additional anti-self-dual two-forms
on ${\cal B}$, that, together with $X^{(1)}$, satisfy the algebra of
unit quaternions,
\begin{equation}
{{X^{(i)}}_m}^p {{X^{(j)}}_p}^n = -\delta^{ij}{\delta_m}^n +
\epsilon^{ijk} {{X^{(k)}}_m}^n\,.
\end{equation}
The coefficient $\lambda^1$ is fixed in terms of the base space
geometry,
\begin{equation}
 \lambda^1 = \frac{\sqrt 3}{\chi R} \Big(\, 
            \frac 12 \nabla^m\nabla_m R + \frac
            23 {\cal R}_{mn} {\cal R}^{mn} - \frac 13 R^2 \Big)\,,
\end{equation}
with $\nabla$ denoting the Levi-Civit\`a connection on the base
manifold with respect to $h$. If we adopt complex coordinates $z^j,
z^{\bar{\jmath}}$ on ${\cal B}$ with respect to $X^{(1)}$
(i.~e.~${{X^{(1)}}^j}_k = i{\delta^j}_k$, ${{X^{(1)}}^{\bar{\jmath}}}_{\bar k}
= -i{\delta^{\bar{\jmath}}}_{\bar k}$),
then $\lambda^2$ and $\lambda^3$ are determined by the differential
equation
\begin{equation}
   \Theta_j = -(\partial_j - iP_j)[R(\lambda^2 - i\lambda^3)]\,,
   \label{equ_lambda}
\end{equation}
which implies that $\lambda^2 - i\lambda^3$ is fixed up to an
arbitrary antiholomorphic function on the base. In (\ref{equ_lambda}),
$P$ and $\Theta$ are given by
\eqn 
P_m &=& \frac 18({X^{(3)}}^{np}\nabla_m {X^{(2)}}_{np} -
               {X^{(2)}}^{np}\nabla_m {X^{(3)}}_{np})\,,
        \label{Pm} \\
\Theta_m &=& {{X^{(2)}}_m}^n (\ast_4 T)_n\,, \label{Theta} 
\feqn
with
\begin{displaymath}
 T = \frac{\sqrt 3}{\chi}\left(-dR\wedge {\cal R} + d\left[\frac 12
\nabla^m\nabla_m R + \frac 23 {\cal R}^{mn}{\cal R}_{mn} -
\frac{1}{12}R^2\right]\wedge X^{(1)}\right)\,.
\end{displaymath}

In summary, $f$ and $G^{\pm}$ are fixed by the geometry of the base
manifold (up to an antiholomorphic function); then, $\omega$ is given
by (\ref{omega}) and finally the gauge potential reads
\begin{equation}
A_m = \chi^{-1}P_m + \frac{\sqrt 3}{2}f\omega_m\,, \qquad A_t = 
\frac{\sqrt 3}2 f\,.
\end{equation}
Note that all fields are independent of $t$.

In order to obtain black hole solutions immersed in a G\"odel-type universe,
we choose as metric on the base manifold
\begin{equation}
h_{mn}dx^m dx^n = H^{-2}dr^2 + \frac{r^2}4H^2(\sigma_3^L)^2 + \frac{r^2}4
                  [(\sigma_1^L)^2 + (\sigma_2^L)^2]\,, \label{basemetr}
\end{equation}
where
\begin{equation}
H(r) = \sqrt{1 + \frac{\chi^2}{12}r^2(1 + \frac{\mu}{r^2})^3}\,.
\end{equation}
For $\mu =0$, this metric reduces to the Bergmann metric, which is
Einstein-K\"ahler. The right-invariant (or "left") one-forms on SU(2)
are given by
\begin{eqnarray}
\sigma_1^L &=& \sin\phi d\theta - \cos\phi\sin\theta d\psi\,, \nonumber \\
\sigma_2^L &=& \cos\phi d\theta + \sin\phi\sin\theta d\psi\,, \nonumber \\
\sigma_3^L &=& d\phi + \cos\theta d\psi\,,
\end{eqnarray}
with the Euler angles $0\le\theta\le\pi$, $0\le\phi\le 2\pi$,
$0\le\psi\le 4\pi$. By introducing the complex coordinates
\begin{equation}
z^1 = h(r)\cos\frac{\theta}2 e^{\frac i2 (\phi + \psi)}\,, \qquad
z^2 = h(r)\sin\frac{\theta}2 e^{\frac i2 (\phi - \psi)}\,,
      \label{complcoord}
\end{equation}
where
\begin{equation}
h(r) = \exp\int\frac{dr}{H^2 r}\,, \label{h}
\end{equation}
one can verify that (\ref{basemetr}) is K\"ahler, with K\"ahler 
potential given by
\begin{equation}
K(r) = \int\frac{r\,dr}{H^2}\,.
\end{equation} 
For the metric (\ref{basemetr}) one finds \cite{Gauntlett:2003fk}:
$\Theta = 0$, $R = -2\chi^2(1 + \mu/r^2)$ and
\begin{eqnarray}
X^{(1)} &=& d(\frac{r^2}4 \sigma_3^L)\,, \nonumber \\
X^{(2)} &=& H^{-1}\frac r2 dr \wedge \sigma_1^L + 
\frac{r^2}4 H d\sigma_1^L\,, \nonumber \\
X^{(3)} &=& H^{-1}\frac r2 dr \wedge \sigma_2^L + \frac{r^2}4 H d\sigma_2^L\,,
\end{eqnarray}
as well as
\begin{equation}
P_{z^1} = \frac{i\chi^2}{8r^2h(r)^2}(r^2 + \mu)^2 \bar{z}^1\,, \qquad
P_{z^2} = \frac{i\chi^2}{8r^2h(r)^2}(r^2 + \mu)^2 \bar{z}^2\,. \qquad
\end{equation}
As $\Theta = 0$, Eq.~(\ref{equ_lambda}) admits the trivial solution
$\lambda^2 = \lambda^3 = 0$ giving the supersymmetric electrically
charged AdS black holes\footnote{Actually these solutions describe
naked singularities. In a slight abuse of notation, we shall
nevertheless refer to them as black holes.} first constructed in
\cite{London:ib}\footnote{For generalizations to the case of gauged
supergravity coupled to vector multiplets see \cite{Behrndt:1998ns}.}.
It is however possible to solve (\ref{equ_lambda})
in general. To this end, we note that
\begin{equation}
P_i = \partial_i L(r)\,,
\end{equation}
where
\begin{equation}
L(r) = \int \frac{i\chi^2}{4H^2r^3}(r^2 + \mu)^2\, dr\,.
\end{equation}
This leads to the general solution
\begin{equation}
\lambda^2 - i\lambda^3 = \frac{{\cal F}(\bar{z}^1, \bar{z}^2)}R e^{iL}\,,
\end{equation}
with ${\cal F}$ denoting an arbitrary antiholomorphic function. If we
choose ${\cal F}$ to be constant, we get the supersymmetric solution
\begin{eqnarray}
ds^2 &=& f^2(dt + \omega)^2 - f^{-1}(H^{-2}dr^2 + \frac{r^2}4 H^2 
(\sigma_3^L)^2
         + \frac{r^2}4 [(\sigma_1^L)^2 + (\sigma_2^L)^2])\,, \nonumber \\
A &=& \frac{\sqrt 3}2f[dt - {\cal F}_1 h^2(r) \sigma_1^L + 
{\cal F}_2 h^2(r) \sigma_2^L]\,,
      \label{goedelBH}
\end{eqnarray}
with
\begin{eqnarray}
f^{-1} &=& 1 + \frac{\mu}{r^2}\,, \nonumber \\
\omega &=& \frac{\chi r^2}{4\sqrt 3}\left(1 + \frac{\mu}{r^2}\right)^3 \sigma_3^L
- {\cal F}_1 h^2(r) \sigma_1^L + {\cal F}_2 h^2(r) \sigma_2^L\,,
\end{eqnarray}
where ${\cal F}_1$ and ${\cal F}_2$ are arbitrary constants related to
the real and imaginary part of ${\cal F}$ respectively and the
function $h(r)$ is given by (\ref{h}).

For $\mu = \chi = 0$, the solution (\ref{goedelBH}) reduces to the
maximally supersymmetric G\"odel-type universe found in
\cite{Gauntlett:2002nw}. Turning on the parameter $\mu$ while
keeping $\chi = 0$ yields the one half supersymmetric G\"odel black hole
studied in detail in \cite{Herdeiro:2002ft}.
For ${\cal F}_1 = {\cal F}_2 = 0$, $\chi \neq 0$, we recover the AdS
black holes of \cite{London:ib}. In the case
$\mu = 0$, $\chi \neq 0$ (\ref{goedelBH}) describes a generalization
of the G\"odel-type universe of \cite{Gauntlett:2002nw} to include a
cosmological constant. This solution was first given in
\cite{Gauntlett:2003fk}, and its chronological structure was studied
in \cite{Behrndt:2003gc}. Although the geometry (\ref{goedelBH}) has
a naked singularity at $r^2 + \mu = 0$ ($R_{\mu\nu\rho\lambda}
R^{\mu\nu\rho\lambda} \sim \frac{f^{11}}{ H^6}$) not hidden by an
event horizon, we shall refer to it as a black hole immersed in a
G\"odel-type universe with cosmological constant\footnote{For
black holes in G\"odel spacetimes without cosmological constant
cf.~\cite{Herdeiro:2002ft,Gimon:2003ms}.}. In section \ref{moregen} we
shall construct G\"odel-type deformations of AdS black holes with
genuine horizons.

\subsection{Physical discussion}

In what follows, we will discuss some physical properties of
(\ref{goedelBH}).  First of all, let us consider its chronological
structure. One finds that the induced metric on hypersurfaces of
constant $t$ and $r$ is always spacelike iff
\begin{equation}
g(r) \equiv {\cal F}_1^2 + {\cal F}_2^2 + 
\sqrt{({\cal F}_1^2 + {\cal F}_2^2)^2
     + \frac{\chi^2 r^4}{12f^6h^4}({\cal F}_1^2 + {\cal F}_2^2)}
     - \frac{r^2}{2f^3h^4} < 0\,.
\end{equation}
For $g(r)>0$ it becomes timelike and thus closed timelike curves
(CTCs) appear.  When we approach the naked singularity at $r^2=-\mu$
(where $f \rightarrow \infty$), $g(r)$ goes to $2({\cal F}_1^2 + {\cal
F}_2^2)$ and thus, as long as ${\cal F}_1$ and ${\cal F}_2$ are
nonvanishing, we have always CTCs near the singularity. On the other
hand, for $r \to \infty$, $g(r)$ is negative provided
\begin{equation}
\frac{\chi^2}3 h^4(\infty)({\cal F}_1^2 + {\cal F}_2^2) < 1\,, \label{bound}
\end{equation}
where $h(\infty)$ indicates the value of $h$ at infinity, which is
easily shown to be a constant. If (\ref{bound}) is satisfied, there
are {\em no} CTCs if $r$ is sufficiently large. However, this conclusion
is only valid on constant time slices, i.e.\ $dt=0$. If we instead allow $t$
to vary, one can construct through every point in spacetime a
CTC. Namely, by going inside the future light cone towards the black
hole singularity, constructing there a time machine as discussed in
\cite{Behrndt:2003gc} and finally coming back to the starting point.
We expect ${\cal F}_1^2 + {\cal F}_2^2$ to measure the angular
momentum/magnetic flux, so (\ref{bound}) should have an interpretation
as a bound on the angular momentum or magnetic flux of the
solution (\ref{goedelBH}). We shall come back to this point below.

Asymptotically for $r\to\infty$ the metric (\ref{goedelBH}) does not
approach AdS$_5$, but a deformation thereof.  The induced metric on
hypersurfaces of constant $r$ is, for large $r$, conformal to
\begin{equation}
ds^2 = \frac{\chi}{2\sqrt 3}\sigma_3^L(dt - {\cal F}_1 h^2(\infty)
       \sigma_1^L + {\cal F}_2 h^2(\infty)\sigma_2^L) - \frac 14
       [(\sigma_1^L)^2 + (\sigma_2^L)^2 + (\sigma_3^L)^2]\,,
       \label{CFTmetric}
\end{equation}
which is always nondegenerate. If ${\cal F}_1$ and ${\cal F}_2$ were
zero, (\ref{CFTmetric}) would be the standard metric on $\bR$ $\times$
S$^3$ (after setting $\phi = \phi' + \chi t/\sqrt 3$), but for ${\cal
F}_1$ or ${\cal F}_2$ different from zero, (\ref{CFTmetric}) describes
a deformation of this standard metric.  According to the AdS/CFT
correspondence, the bulk solution (\ref{goedelBH}) should have a dual
description in terms of (a deformation of) $D=4$, ${\cal N}=4$
super-Yang-Mills theory defined on the curved manifold (\ref{CFTmetric}).
In what follows we shall consider more in detail the supergravity solution
(\ref{goedelBH}) with $\mu = 0$, i.~e.~, the G\"odel-deformation of AdS$_5$,
and its CFT dual. We have then
\begin{displaymath}
h^2(r) = \frac{C r^2}{1 + \frac{\chi^2}{12}r^2}\,,
\end{displaymath}
where $C$ denotes an arbitrary integration constant that can be absorbed
into ${\cal F}_{1,2}$. In order to see which operators/VEVs are turned on
in the CFT, one has to do a Fefferman-Graham expansion of the supergravity
fields\footnote{For a nice review of the procedure see \cite{Skenderis:2002wp}.},
which in our case consist of the metric and the U(1) gauge field
only. For the gauge field we have the asymptotic behaviour
\begin{equation}
A = \frac{6\sqrt 3}{\chi^2}[-{\cal F}_1 \sigma_1^L + {\cal F}_2 \sigma_2^L]
    (1 - 12 \chi^{-2}r^{-2} + {\cal O}(r^{-4}))\,. \label{FeffGrahamA}
\end{equation}
Now a massless vector field $A$ in AdS$_5$, which naturally couples to
a CFT R-current $J$, typically falls off like
$r^{-2}$ or like $r^0$ for $r \to \infty$\footnote{A gauge invariant way of saying this
is that the "electric" field $F_{r a}$ ($a = t, \phi, \theta, \psi$) falls off like
$r^{-3} + {\cal O}(r^{-5})$ and the "magnetic" field $F_{ab}$ like
$r^0 + {\cal O}(r^{-2})$.}.
The latter behaviour is the non-normalizable mode corresponding to the insertion of
the dual operator. From (\ref{FeffGrahamA}) we see that in our case the dual
operator is inserted,
i.~e.~, the CFT is deformed by the term
\begin{equation}
\int d^4 x \sqrt{-\gamma} A^I_{\mu} J^{\mu}_I\,, \label{AJ}
\end{equation}
where $\gamma$ denotes the determinant of the metric (\ref{CFTmetric})
and $I$ is an SO(6) R-symmetry index. (In our case $A^I$ takes values
in the Cartan subgroup SO(2) $\times$ SO(2) $\times$ SO(2) of SO(6),
with all three components equal). $J$ has dimension $\Delta = 3$,
and thus our bulk solution is described by a relevant deformation of
${\cal N}=4$ super-Yang-Mills theory residing on the curved
manifold (\ref{CFTmetric}). Of course this deformation preserves
only part of the original supersymmetry.
The situation encountered here is somewhat
similar to that for the G\"odel black hole without cosmological
constant studied in \cite{Herdeiro:2002ft}. Both this spacetime
and the BMPV black hole are described (after uplifting to ten dimensions)
by a deformation of the D1-D5-pp-wave system, but the BMPV perturbation
is normalizable whereas the G\"odel perturbation is non-normalizable
and corresponds to the insertion of an operator in the dual two-dimensional
CFT \cite{Herdeiro:2002ft}. (In the BMPV case, symmetry is broken spontaneously,
whereas in the G\"odel case it is broken explicitely).
For the BMPV black hole, the rotation
corresponds to a VEV of the CFT R-currents. Now the
classification of unitary representations of superconformal algebras
typically yields inequalities on the conformal weights and R-charges.
Generically unitarity is violated if the R-charges become too large.
It has been shown in \cite{Herdeiro:2000ap} that the threshold where
CTCs develop in the bulk of the BMPV black hole (when the angular
momentum becomes too large, overrotating case) corresponds exactly to
a unitarity bound in the dual CFT.

It would be interesting to see whether a similar holographic interpretation
can be given to the bound (\ref{bound}), i.~e.~, if the appearance
of closed timelike curves in the bulk is related to loss of unitarity
in the dual field theory. From (\ref{FeffGrahamA}) we see that $J$ gets
also a VEV. This is of course a consequence of the deformation (\ref{AJ}) of the
CFT Lagrangian, in the same way in which an external magnetic field applied
to a ferromagnet implies a magnetization. The R-charge in our case is given in
terms of the constants ${\cal F}_{1,2}$, so in principle a unitarity bound on
the R-charges could lead to an inequality like (\ref{bound}). To finally answer this
question one needs the residual superalgebra preserved by the solution (\ref{goedelBH})
after lifting to ten dimensions, which we will not attempt to determine here.

In any case, by analyzing (\ref{CFTmetric}), it is straightforward
to show that beyond the bound (\ref{bound}),
the CFT metric itself develops CTCs. This can be seen by considering e.~g.~the
vector $v = \xi_1^L + \xi_3^L$, where
\begin{eqnarray}
\xi^L_1 &=& \sin\phi \partial_{\theta} + \cot\theta \cos\phi \partial_{\phi}
            - \frac{\cos\phi}{\sin\theta}\partial_{\psi}\,, \nonumber \\
\xi^L_2 &=& \cos\phi \partial_{\theta} - \cot\theta \sin\phi \partial_{\phi}
            + \frac{\sin\phi}{\sin\theta}\partial_{\psi}\,, \nonumber \\
\xi^L_3 &=& \partial_{\phi}\,, \nonumber
\end{eqnarray}
denote the left vector fields on SU(2). $v$ has closed orbits and becomes timelike
whenever (\ref{bound}) is violated. We probably cannot make sense of a quantum field
theory on a spacetime with CTCs, e.~g.~the Cauchy problem is ill-defined, there is no
notion of an S-matrix, and so on. This means that beyond the bound (\ref{bound}) the
boundary CFT is probably itself pathological. It seems thus to be a general pattern that
whenever CTCs develop in the bulk, the holographic dual is not well-defined.

It would be desirable to have a mechanism which avoids or forbids these G\"odel-type
deformations that suffer from CTCs. In \cite{Dyson:2003zn} it was argued (for the
example of the overrotating three-charge black hole in five dimensions) that
stringy effects would prohibit any attempt to build the causality violating
regions, i.~e.~, once all stringy effects are taken into account, our usual
notion of chronology will emerge as a protected law of nature. It would be
interesting to see exactly how such a "stringy protection of chronology"
is realized in our case. Another interesting way could be to construct
the non-extreme solutions and to investigate the thermodynamical
stability, but we leave a more detailed discussion of this for future
work.

As a last point of the physical discussion, we compute the holographic
stress tensor \cite{Balasubramanian:1999re} of the solution
(\ref{goedelBH}) with $\mu = 0$\footnote{While this paper has been
revised, ref.~\cite{GGS} appeared in which the holographic
energy-momentum tensor was also calculated.}. To this end, we first write the
metric in a way in which it is manifestly asymptotically AdS (modulo
the deformation mentioned above). After shifting $\phi \to \phi + \chi
t/\sqrt 3$, $t \to t - t_0$, where $\tan(\chi t_0/\sqrt 3) = - {\cal
F}_2/{\cal F}_1$, we get
\begin{eqnarray}
ds^2 &=& \left(1 + \frac{\chi^2}{12}r^2\right)(dt + \tilde{\omega})^2 -
         \frac{dr^2}{1 + \frac{\chi^2}{12}r^2} - \frac{r^2}4 [(\sigma_1^L)^2 
+ (\sigma_2^L)^2
         + (\sigma_3^L - \frac{\chi}{\sqrt 3}\tilde{\omega})^2]\,, \nonumber \\
A &=& \frac{\sqrt 3}2 \tilde{\omega}\,, \label{metrtimedep}
\end{eqnarray}
where we defined
\begin{displaymath}
\tilde{\omega} = \frac{r^2 {\cal F}}{1 + \frac{\chi^2}{12}r^2}\,
\left[\sigma_1^L
\cos\frac{\chi t}{\sqrt 3} + \sigma_2^L\sin\frac{\chi t}{\sqrt 3}\right]\,,
\end{displaymath}
and ${\cal F} = \sqrt{{\cal F}_1^2 + {\cal F}_2^2}$. Note that the
shift of the angle $\phi$ leads to an explicit time-dependence of the
boundary metric.\\ The complete (Lorentzian) action reads
\begin{equation}
S = S_{bulk} + S_{surf} + S_{ct}\,,
\end{equation}
where $S_{bulk}$ is given in (\ref{action}),
\begin{equation}
S_{surf} = -\frac{1}{8\pi G}\int_{\partial{\cal M}} d^4 x 
\sqrt{-\sigma} K \label{GibbHawk}
\end{equation}
denotes the Gibbons-Hawking boundary term required to have a
well-defined variational principle, and
\begin{equation}
S_{ct} = \frac{1}{8\pi G}\int_{\partial{\cal M}} d^4 x \sqrt{-\sigma}
         \left[-\frac{3}{\ell} + \frac{\ell\,{\cal R}}4\right] \label{counter}
\end{equation}
is a surface counterterm introduced in \cite{Balasubramanian:1999re}
to render the total action finite. In (\ref{GibbHawk}) and
(\ref{counter}), $\sigma_{ab}$ is the induced metric on the boundary
$\partial{\cal M}$ of the spacetime ${\cal M}$, and $K$ denotes the
trace of the extrinsic curvature $K_{ab} = -\frac 12 (\nabla_a n_b +
\nabla_b n_a)$ of $\partial{\cal M}$, where $n_a$ is the outward
pointing unit normal to $\partial{\cal M}$.  ${\cal R}$ is the scalar
curvature of $\sigma_{ab}$, and $\ell = 2\sqrt 3/\chi$.  Note that in
five dimensions, in general one encounters also logarithmic
divergences in the computation of the action. These divergences, which
cannot be removed by adding local counterterms like (\ref{counter}),
are related to the Weyl anomaly of the dual CFT
\cite{Henningson:1998gx}. However, as one readily verifies, in our
case there are no logarithmic divergences, so there is no conformal
anomaly in the dual field theory.  We will come back to this point
below.\\ One can now construct a divergence-free stress tensor given
by \cite{Balasubramanian:1999re}
\begin{equation}
T_{ab} = \frac{2}{\sqrt{-\sigma}}\frac{\delta S}{\delta \sigma^{ab}} =
             -\frac{1}{8\pi G}\left[K_{ab} - K\sigma_{ab} -\frac{3}{\ell}
             \sigma_{ab} - \frac{\ell}2 {\cal G}_{ab}\right]\,,
\end{equation}
where ${\cal G}_{ab}$ denotes the Einstein tensor built from $\sigma_{ab}$.
If we choose $\partial{\cal M}$ to be a four-surface of fixed $r$, we get
\begin{eqnarray}
8\pi G T_{tt} &=& \frac{3\ell}{8r^2} + 
\frac{8\ell^3 {\cal F}^2}{r^2} + {\cal O}(r^{-4})\,,
                  \nonumber \\
8\pi G T_{t\phi} &=& 
\frac{4\ell^4{\cal F}^2}{r^2} + {\cal O}(r^{-4})\,, \nonumber \\
8\pi G T_{t\theta} &=& -\frac{\ell^3{\cal F}}{8 r^2}\sin(\phi + 2t/\ell) +
                       {\cal O}(r^{-4})\,, \nonumber \\
8\pi G T_{t\psi} &=& \frac{\ell^3{\cal F}}{8 r^2}
[\sin\theta\cos(\phi + 2t/\ell) +
     32\ell {\cal F}\cos\theta] + {\cal O}(r^{-4})\,, \nonumber \\
8\pi G T_{\phi\phi} &=& \frac{\ell^3}{32 r^2} + 
\frac{2\ell^5 {\cal F}^2}{r^2} + {\cal O}(r^{-4})\,,
                        \nonumber \\
8\pi G T_{\phi\theta} &=& 
-\frac{13\ell^4{\cal F}}{16 r^2}\sin(\phi + 2t/\ell) +
                          {\cal O}(r^{-4})\,, \nonumber \\
8\pi G T_{\phi\psi} &=& 
\frac{\ell^3}{16 r^2}[13\ell{\cal F}\sin\theta\cos(\phi + 2t/\ell) +
                        \frac 12 \cos\theta + 32\ell^2 {\cal F}^2\cos\theta]
                        + {\cal O}(r^{-4})\,, \nonumber \\
8\pi G T_{\theta\theta} &=& \frac{\ell^3}{32 r^2} + 
\frac{5\ell^5 {\cal F}^2}{2 r^2}
     \sin^2(\phi + 2t/\ell) + {\cal O}(r^{-4})\,, \nonumber \\
8\pi G T_{\theta\psi} &=& -\frac{\ell^4 {\cal F}}{4 r^2}
[5\ell\,{\cal F}\sin\theta
          \sin(2\phi + 4t/\ell) + \frac{13}4 \cos\theta\sin(\phi + 2t/\ell)]
                          + {\cal O}(r^{-4})\,, \nonumber \\
8\pi G T_{\psi\psi} &=& \frac{\ell^3}{8 r^2}
\left[\frac 14 + 20\ell^2 {\cal F}^2\sin^2\theta
  \cos^2(\phi + 2t/\ell) + 16\ell^2 {\cal F}^2\cos^2\theta\right. \nonumber \\
            & & \qquad \left. + 
13\ell\,{\cal F}\sin\theta\cos\theta\cos(\phi + 2t/\ell)\right]
                        + {\cal O}(r^{-4})\,. \nonumber
\end{eqnarray}
The metric on the manifold on which the dual CFT resides is defined by
\begin{displaymath}
\gamma_{ab} = \lim_{r\to\infty}\frac{\ell^2}{r^2}\sigma_{ab}\,,
\end{displaymath}
which yields
\begin{equation}
\gamma_{ab}\,dx^a dx^b = (dt + \Omega)^2 - \frac{\ell^2}4
[(\sigma_1^L)^2 + (\sigma_2^L)^2 + (\sigma_3^L - \frac{2}{\ell}\Omega)^2]\,,
\end{equation}
where
\begin{displaymath}
\Omega = \ell^2 {\cal F}\,\left(\sigma_1^L\cos\frac{2t}{\ell} + 
\sigma_2^L\sin\frac{2t}{\ell}\right)\,.
\end{displaymath}
The field theory stress tensor $\hat{T}^{ab}$ is related to $T^{ab}$ by
the rescaling \cite{Myers:1999ps}
\begin{displaymath}
\sqrt{-\gamma}\,\gamma_{ab}\hat{T}^{bc} = \lim_{r \to \infty}
\sqrt{-\sigma}\,\sigma_{ab}T^{bc}\,,
\end{displaymath}
which amounts to multiplying all expressions for $T_{ab}$ given above
by $r^2/\ell^2$ before taking the limit $r \to \infty$, in order to
obtain $\hat{T}_{ab}$. Alternatively, the CFT energy-momentum tensor
$\hat{T}_{ab}$ could have been obtained from a Fefferman-Graham
expansion of the five-dimensional metric
\cite{Myers:1999ps,Skenderis:2002wp}. We see that in our case, apart
from the R-current, also the stress tensor gets a VEV. We have checked
that $\hat{T}_{ab}$ is conserved and traceless, ${\cal D}_a
\hat{T}^{ab} = 0$, $\gamma^{ab}\hat{T}_{ab} = 0$, where ${\cal D}$
denotes the connection of the metric $\gamma$. The tracelessness means
that there is no conformal anomaly in the dual CFT.\\ The holographic
stress tensor can also be used to compute conserved quantities like
mass and angular momentum of the spacetime. To do this, we indicate by
$u^{\mu}$ the unit normal vector of a spacelike hypersurface $^4{\cal
S}_t$ at constant $t$, and by $\Sigma$ the spacelike intersection
$^4{\cal S}_t \cap \partial {\cal M}$ embedded in $\partial {\cal M}$
with induced metric $\Sigma_{ab}$. Then, for any Killing vector field
$\xi^{\mu}$ there is an associated conserved charge
\begin{equation}
Q_{\xi} = \int_{\Sigma} d^3x\,\sqrt{-\Sigma}\,u^{\mu}T_{\mu\nu}\xi^{\nu}\,.
          \label{Qxi}
\end{equation}
Under the shift $\phi \to \phi + 2t/\ell$, the Killing vector $\partial_t$ goes
to $\partial_t - \frac{2}{\ell}\partial_{\phi}$. We find that the conserved
charge associated to this Killing vector, which we will call the mass of the
solution, is given by
\begin{equation}
M = \frac{3\pi\ell^2}{32 G} + \frac{\pi \ell^4 {\cal F}^2}{G}\,.
\end{equation}
Using the AdS/CFT dictionary $\ell^3/G = 2N^2/\pi$, this can also be written as
\begin{equation}
M = \frac{3N^2}{16 \ell} + 2N^2 \ell\,{\cal F}^2\,.
\end{equation}
The first term is just the Casimir energy for ${\cal N} = 4$ SYM on
$\bR \times S^3$ \cite{Balasubramanian:1999re}, whereas the second
term arises from the G\"odel-deformation.  The isometry group of the
G\"odel-AdS$_5$ spacetime (\ref{goedelBH}) (with $\mu = 0$) is $\bR
\times$ SU(2)$_R$, where $\bR$ is generated by $\partial_t -
\frac{2}{\ell}\partial_{\phi}$, and SU(2)$_R$ by the right vector
fields (\ref{rightvec}). The conserved charges associated to $\xi_i^R$
turn out to be zero, so the angular momenta of the solution vanish. At
first sight, this might seem surprising, because the one-form $\omega$
in (\ref{goedelBH}) causes a rotation. The point is that the amount of
this rotation depends on the angles on the three-sphere, so locally
the spacetime rotates, but globally not. (The angular momentum density
appearing in (\ref{Qxi}) is nonvanishing, but the integral over it is
zero). Although there is no {\it global} angular momentum, the
additional term appearing in the mass $M$ proportional to ${\cal F}^2$
might have an interpretation as rotational energy, because the
associated energy density is proportional to the square of the angular
momentum density.

\section{More general supersymmetric solutions}

\label{moregen}

The solution (\ref{goedelBH}) is actually a special case of a more general
class of BPS solutions, that are given in terms of a base space ${\cal B}$
which is a complex line bundle over a two-dimensional surface $\Sigma$.
For the metric on ${\cal B}$ we choose
\begin{equation}
h_{mn}dx^m dx^n = \frac{dr^2}{V(r)} + V(r)(d\phi + 
{\cal A})^2 + F^2(r)d\Sigma^2\,,
       \label{linebundle}
\end{equation}
where the one-form ${\cal A}$ on $\Sigma$ and the functions $V(r), F(r)$ will
be determined below by requiring that (\ref{linebundle}) be the metric on a
K\"ahler manifold.
Although arbitrary surfaces $\Sigma$
might be possible, we shall consider only the case where $\Sigma$ is
a space of constant curvature $k$, where without loss of generality
$k=0,\pm 1$. For the line element $d\Sigma^2$ on $\Sigma$ we can take
\begin{equation}
d\Sigma^2 = d\theta^2 + S^2(\theta)d\psi^2\,,
\end{equation}
with
\begin{equation}
S(\theta) = \left\{ \begin{array}{r@{\quad,\quad}l}
                   \sin\theta & k=1\,, \\ \sinh\theta & k=-1\,, \\ 1 & k=0\,.
                   \end{array} \right.
\end{equation}
The anti-self-dual two-forms $X^{(i)}$ on the base manifold ${\cal B}$ can be
chosen as
\begin{eqnarray*}
X^{(1)} &=& e^1 \wedge e^2 - e^3 \wedge e^4\,, \\
X^{(2)} &=& e^1 \wedge e^3 + e^2 \wedge e^4\,, \\
X^{(3)} &=& e^1 \wedge e^4 - e^2 \wedge e^3\,,
\end{eqnarray*}
where the vierbein is given by
\begin{displaymath}
e^1 = \frac{dr}{\sqrt{V(r)}}\,, \quad e^2 = \sqrt{V(r)}\sigma_3\,, \quad
e^3 = F(r)\sigma_1\,, \quad e^4 = F(r)\sigma_2\,,
\end{displaymath}
and we defined
\begin{eqnarray}
\sigma_1 &=& \sin\alpha\phi\, d\theta - 
S(\theta)\cos\alpha\phi\, d\psi\,, \nonumber \\
\sigma_2 &=& \cos\alpha\phi\, d\theta + 
S(\theta)\sin\alpha\phi\, d\psi\,, \label{sigma} \\
\sigma_3 &=& d\phi + {\cal A}\,, \nonumber
\end{eqnarray}
with $\alpha$ to be determined below. $X^{(1)}$ is then closed provided
\begin{equation}
{\cal A} = \left\{ \begin{array}{r@{\quad,\quad}l}
           n\cos\theta\, d\psi & k=1\,, \\ -n\cosh\theta\, d\psi & k=-1\,, \\
                   \frac n2 (\psi\, d\theta - \theta\, d\psi) & k=0\,,
                   \end{array} \right. \label{A=Kaehler}
\end{equation}
$\alpha = k/n$ and
\begin{equation}
F^2(r) = nr\,, \label{F^2}
\end{equation}
where $n$ is an arbitrary constant. Eq.~(\ref{A=Kaehler}) means that
$d{\cal A}$ is proportional to the K\"ahler form on $\Sigma$. It can
then be checked that for arbitrary $V(r)$, the two-forms $X^{(2)}$ and
$X^{(3)}$ satisfy
\begin{eqnarray}
\nabla_m X^{(2)}_{np} &=& P_m X^{(3)}_{np}\,, \nonumber \\
\nabla_m X^{(3)}_{np} &=& -P_m X^{(2)}_{np}\,, \label{covdiffX}
\end{eqnarray}
with $P$ given by
\begin{equation}
P = \left(\frac kn - \frac{V'(r)}2 - \frac{V(r)}{2r}\right)\sigma_3\,.
\end{equation}
Note that (\ref{covdiffX}) implies (\ref{Pm}).

In conclusion, (\ref{linebundle}) is a K\"ahler metric for arbitrary
function $V(r)$, provided ${\cal A}$ and $F(r)$ satisfy
(\ref{A=Kaehler}) and (\ref{F^2}) respectively. The K\"ahler form on
${\cal B}$ is given by $X^{(1)}$. This general base manifold ${\cal
B}$ can be used as a starting point for the construction of a variety
of new supersymmetric solutions of minimal gauged supergravity in five
dimensions. Note that for general $V(r)$, the one-form $\Theta$
defined in (\ref{Theta}) does not vanish, which makes it rather
difficult to solve equation (\ref{equ_lambda}). If we choose $n=1$ and
\begin{equation}
V(r) = r\left[k + \frac{\chi^2}3 r\left(1 + \frac{\mu}r\right)^3\right]\,,
       \label{V}
\end{equation}
where $\mu$ denotes an arbitrary parameter, $\Theta$ vanishes.
The base manifold has then the scalar curvature
\begin{equation}
R = -2\chi^2\left(1 + \frac{\mu}r\right)\,,
\end{equation}
which yields for the function $f$
\begin{equation}
f^{-1} = 1 + \frac{\mu}{r}\,.
\end{equation}
The spherical case $k=1$ leads (after the coordinate transformation $r
\to r^2/4$) to the supersymmetric G\"odel black hole (\ref{goedelBH})
already discussed above.  Let us therfore focus our attention to the
cases $k=-1$ and $k=0$.  For $k=-1$, $\Sigma$ is a hyperbolic space
(or a quotient thereof). One can choose the complex coordinates
\begin{equation}
z^1 = h(r)\cosh\frac{\theta}2 e^{-\frac i2(\phi - \psi)}\,, \qquad
z^2 = h(r)\sinh\frac{\theta}2 e^{-\frac i2(\phi + \psi)}\,,
\end{equation}
on ${\cal B}$ with respect to $X^{(1)}$, where
\begin{equation}
h(r) = \exp\left[-\int\frac{dr}{2V(r)}\right]\,. \label{hk=-1}
\end{equation}
This leads to
\begin{equation}
P_{z^1} = -\frac{i\chi^2}{2rh(r)^2}(r + \mu)^2 \bar{z}^1\,, \qquad
P_{z^2} = \frac{i\chi^2}{2rh(r)^2}(r + \mu)^2 \bar{z}^2\,, \qquad
\end{equation}
or
\begin{equation}
P_i = \partial_i L(r)\,, \qquad L(r) = \int\frac{i\chi^2}{2Vr}(r+\mu)^2\,dr
      \label{P=dL}
\end{equation}
for the holomorphic components of the one-form $P$. Using this as well
as $\Theta=0$, one can solve Eq.~(\ref{equ_lambda}) to obtain
$\lambda^2$ and $\lambda^3$ up to an arbitrary antiholomorphic
function ${\cal F}(\bar{z}^1, \bar{z}^2)$, which we will take to be
constant. One arrives then finally at the BPS solution
\begin{eqnarray}
ds^2 &=& f^2(dt + \omega)^2 - f^{-1}(V^{-1}dr^2 + V (\sigma_3)^2
         + r [(\sigma_1)^2 + (\sigma_2)^2])\,, \nonumber \\
A &=& \frac{\sqrt 3}2f[dt - {\cal F}_1 h^2(r) \sigma_1 + 
{\cal F}_2 h^2(r) \sigma_2]\,,
      \label{goedelBHk=-1}
\end{eqnarray}
with
\begin{equation}
\omega = \frac{\chi r}{\sqrt 3}f^{-3} \sigma_3
- {\cal F}_1 h^2(r) \sigma_1 + {\cal F}_2 h^2(r) \sigma_2\,,
\end{equation}
where ${\cal F}_1$ and ${\cal F}_2$ are arbitrary constants related to
the real and imaginary part of the antiholomorphic function ${\cal
F}$.  The one-forms $\sigma_i$ and $h(r)$ are given by (\ref{sigma})
and (\ref{hk=-1}) respectively and for $\mu=0$, (\ref{goedelBHk=-1})
reduces to the solution found in \cite{Behrndt:2003gc}. For ${\cal
F}_1 = {\cal F}_2 = 0$, we recover the hyperbolic black holes of
\cite{Behrndt:1998jd}.

For $k=0$, $\Sigma$ is flat and as in \cite{Behrndt:2001km}, we choose
as complex coordinates
\begin{equation}
\zeta = \frac 12 (\theta - i\psi)\,, \qquad
S = -\int\frac{dr}{V(r)} - i\phi + \frac 14 (\theta^2 + \psi^2)\,,
\end{equation}
in terms of which the base space metric reads
\begin{equation}
h_{mn}dx^m dx^n = V(dS - 2\bar{\zeta}\,d\zeta)
   (d\bar S - 2\zeta\,d\bar{\zeta})
                  +4r\,d\zeta\,d\bar{\zeta}\,.
\end{equation}
This yields
\begin{equation}
P_S = -\frac{i\chi^2}{4r}(r + \mu)^2\,, \qquad
P_{\zeta} = \frac{i\chi^2}{2r}(r + \mu)^2\bar{\zeta}\,,
\end{equation}
which implies again (\ref{P=dL}). Proceeding like in the cases $k=\pm 1$, we
get then the supersymmetric solution
\begin{eqnarray}
ds^2 &=& f^2(dt + \omega)^2 - f^{-1}(V^{-1}dr^2 + V (\sigma_3)^2
         + r [(\sigma_1)^2 + (\sigma_2)^2])\,, \nonumber \\
A &=& \frac{\sqrt 3}2f[dt - {\cal F}_1(-h^2(r)\,d\psi + 
  \phi\,d\theta + \frac 14
      \theta^2\,d\psi) \nonumber \\
  & & \qquad + {\cal F}_2(h^2(r)\,d\theta + 
      \phi\,d\psi - \frac 14\psi^2\,d\theta)]\,,
      \label{goedelBHk=0}
\end{eqnarray}
with
\begin{eqnarray*}
\omega &=& \frac{\chi r}{\sqrt 3}f^{-3} \sigma_3 - {\cal F}_1(-h^2(r)\,
    d\psi + \phi\,d\theta + \frac 14 \theta^2\,d\psi) \\
       & & \qquad + {\cal F}_2(h^2(r)\,d\theta +
           \phi\,d\psi - \frac 14\psi^2\,d\theta)\,,
\end{eqnarray*}
where again ${\cal F}_1$, ${\cal F}_2$ are constants,
\begin{displaymath}
h^2(r) = \int\frac{dr}{V(r)}\,,
\end{displaymath}
and the one-forms $\sigma_i$ are given by (\ref{sigma}).  For $\mu=0$,
(\ref{goedelBHk=0}) reduces to the solution found in
\cite{Behrndt:2003gc} whereas for ${\cal F}_1 = {\cal F}_2 = 0$, we
obtain the supersymmetric black holes of \cite{Behrndt:1998jd}. It is
interesting to note that these black holes were recovered in
\cite{Gauntlett:2003fk} by taking a different base manifold. This
means that different base geometries can lead to the same BPS solution
of gauged supergravity.

As a final choice, which includes both (\ref{V}) and the
supersymmetric AdS$_5$ black holes obtained recently in
\cite{Gutowski:2004ez}, we take $k=1$, $n=1$ and
\begin{equation}
V(r) = a_2 r^2 + a_1 r + a_0 + \frac{a_{-1}}r\,,
\end{equation}
which behaves for large $r$ as the Bergmann metric and for small
$r$ as the black hole discussed before. But depending on the parameters
there can be a horizon for some finite $r$; see below. As the only
restriction on the parameters, we impose $\Theta = 0$ yielding
\begin{equation}
3a_{-1}(a_1 - 1) = a_0^2\,,
\end{equation}
and hence the base space is now parameterized by three parameters.
The scalar curvature of the base space becomes then
\begin{equation}
R = -2\left(3a_2 + \frac{a_1-1}r\right)\,,
\end{equation}
which implies
\begin{equation}
f^{-1} = \frac{3a_2}{\chi^2} + \frac{a_1-1}{\chi^2 r}\,.
\end{equation}
In order to solve (\ref{equ_lambda}), we introduce complex coordinates
as in (\ref{complcoord}), with $h(r)$ given by
\begin{equation}
h(r) = \exp\int\frac{dr}{2V(r)}\,.
\end{equation}
This leads to
\begin{equation}
P_{z^1} = -\frac{i}{h^2}\left(1 - \frac{V'}2 - 
\frac V{2r}\right)\bar{z}^1\,, \qquad
P_{z^2} = -\frac{i}{h^2}\left(1 - \frac{V'}2 - 
\frac V{2r}\right)\bar{z}^2\,,
\end{equation}
or
\begin{equation}
P_i = \partial_i L(r)\,, \qquad L(r) = 
   -i\int\left(1 - \frac{V'}2 - \frac V{2r}\right)
      \frac{dr}V
\end{equation}
for the holomorphic components of the one-form $P$. Using this as well
as $\Theta = 0$, one can again solve Eq.~(\ref{equ_lambda}) to obtain
$\lambda^2$ and $\lambda^3$ up to an arbitrary antiholomorphic
function ${\cal F}(\bar{z}^1,\bar{z}^2)$, which we take as usual to be
constant. One obtains then the supersymmetric solution
\begin{eqnarray}
ds^2 &=& f^2(dt + \omega)^2 - f^{-1}(V^{-1}dr^2 + V (\sigma_3^L)^2
         + r [(\sigma_1^L)^2 + (\sigma_2^L)^2])\,, \nonumber \\
A &=& \frac{\sqrt 3}2f[dt - {\cal F}_1 h^2(r) \sigma_1^L + 
  {\cal F}_2 h^2(r) \sigma_2^L]
      \nonumber \\
  & & + \frac{3a_0 a_2 - (a_1-1)^2}{4\chi^3 r}f \sigma_3^L\,,
      \label{goedelBHk=1gen}
\end{eqnarray}
with
\begin{eqnarray}
\omega &=& \frac{3r(a_1-1)^2 + (18 a_2 r^2 + 2a_0)
  (a_1-1) + 18 r^3 a_2^2 + 9a_0 a_2 r}
           {2\sqrt 3 r^2 \chi^3} \sigma_3^L \nonumber \\
     & & - {\cal F}_1 h^2(r) \sigma_1^L + {\cal F}_2 h^2(r) \sigma_2^L\,. 
	   \nonumber
\end{eqnarray}
Note that by rescaling
\begin{displaymath}
t \to \gamma^{-1} t\,, \quad r \to \gamma r\,, \quad a_0 \to \gamma a_0\,, 
\quad
a_2 \to \gamma^{-1} a_2\,, \quad {\cal F}_{1,2} \to 
\gamma^{-1}{\cal F}_{1,2}\,,
\end{displaymath}
we can set $a_2 = \chi^2/3$.

We recover our solution (\ref{goedelBH}) if $a_{-1} = \mu^3
\chi^2/3$, $a_0 = \mu^2 \chi^2$, $a_1 = 1 + \mu \chi^2$, whereas the
choice $a_{-1} = a_0 = 0$, $a_1 = 4a^2$, ${\cal F}_{1,2} = 0$ yields
the rotating supersymmetric black holes with regular event horizon
obtained recently in \cite{Gutowski:2004ez}. (The rotation parameter
$a$ corresponds to their $\alpha$; the radial coordinate $\rho$ used
in \cite{Gutowski:2004ez} is related to $r$ by
$r = 12 a^2 \chi^{-2}\sinh^2\frac{\chi \rho}{2\sqrt 3}$.)
As before, the G\"odel deformation for this black hole
corresponds to non-vanishing values of ${\cal F}_{1,2}$, which gives
\begin{eqnarray}
ds^2 &=& f^2(dt + \omega)^2 - f^{-1}\Big( V^{-1}dr^2 + V (\sigma_3^L)^2
         + r \big[(\sigma_1^L)^2 + (\sigma_2^L)^2\big] \Big)\,, 
	 \nonumber \\[2mm]
A &=& \frac{\sqrt 3}2f\Big[ dt + 
  \Big(1 + \frac{12 a^2}{\chi^2 r}\Big)^{-\frac 1{4a^2}}
  \Big(-{\cal F}_1 \sigma_1^L
      + {\cal F}_2 \sigma_2^L\Big) \Big] - \frac{(4a^2-1)^2}{4\chi^3 r}f 
\sigma_3^L\,, \nonumber
\end{eqnarray}
with
\begin{displaymath}
\omega = \frac{3(4a^2 - 1)^2 + 6\chi^2 r (4a^2 - 1) + 2 r^2 \chi^4}
       {2\sqrt 3 r \, \chi^3}
         \sigma_3^L + \Big(1 + \frac{12 a^2}{\chi^2 r}\Big)^{-\frac 1{4a^2}}
	 \Big(-{\cal F}_1 \sigma_1^L
         + {\cal F}_2 \sigma_2^L\Big)\,,
\end{displaymath}

\begin{displaymath}
V = \frac{\chi^2}3 r^2 + 4a^2 r\,, \qquad f^{-1} = 
1 + \frac{4a^2 - 1}{\chi^2 r}\,.
\end{displaymath}

Generically this solution contains CTCs. This follows from the fact that
asymptotically for $r \to \infty$ it approaches the G\"odel-type
deformation of AdS$_5$ studied in \cite{Behrndt:2003gc}. One can show
(by expanding $g_{\psi\psi}$ for $r \to \infty$) that
e.~g.~$\partial_{\psi}$ can become timelike (at least for large $r$),
provided
\begin{equation}
{\cal F}_1^2 + {\cal F}_2^2 > \frac 3{\chi^2}\,. 
\end{equation}
It would be nice to see whether the spacetime contains no CTCs at all if
${\cal F}_1^2 + {\cal F}_2^2$ lies below this bound (as is the case
for $a^2 = 1/4$ \cite{Behrndt:2003gc}). We will not attempt to do
this here. Note that $r=0$ is a Killing horizon of the Killing
vector $\xi = \partial_t$. It is straightforward to show that the
surface gravity
\begin{displaymath}
\kappa^2 = -\frac 12 \nabla^{\mu}\xi^{\nu}
\nabla_{\mu}\xi_{\nu}|_{\mathrm{Hor.}}
\end{displaymath}
vanishes, as it must be for supersymmetric black holes.
The isometry group $\bR \times U(1)_L \times SU(2)_R$ of the spacetime
with ${\cal F}_{1,2} = 0$ \cite{Gutowski:2004ez} is broken down to
$\bR \times SU(2)_R$ by the G\"odel deformation, i.~e.~, by nonvanishing
${\cal F}_1$ or ${\cal F}_2$. The $SU(2)_R$ is generated by the right
vector fields
\begin{eqnarray}
\xi^R_1 &=& -\sin\psi \partial_{\theta} - \cot\theta \cos\psi \partial_{\psi}
            + \frac{\cos\psi}{\sin\theta}\partial_{\phi}\,, \nonumber \\
\xi^R_2 &=& \cos\psi \partial_{\theta} - \cot\theta \sin\psi \partial_{\psi}
            + \frac{\sin\psi}{\sin\theta}\partial_{\phi}\,, \label{rightvec} \\
\xi^R_3 &=& \partial_{\psi}\,. \nonumber
\end{eqnarray}
For ${\cal F}_{1,2} = 0$, one has an additional left Killing vector
$\partial_{\phi}$ corresponding to $U(1)_L$.

The general solution (\ref{goedelBHk=1gen}) might have "event
horizons" at $r = r_H > 0$, if $V(r_H) = 0$. It is straightforward to
show that in this case the metric on the horizon would not be
Euclidean, or, in other words, CTCs would develop outside the
horizon. Thus, strictly speaking, the zeroes of $V(r)$ are not really
horizons, rather we expect them to show a repulson-like behaviour, as
in the case of the overrotating BMPV black hole
\cite{Gibbons:1999uv}. (Note that the surface $r = r_H$, $dt = 0$,
which becomes Lorentzian, is tangent to what we would naively call a
horizon. This means that there are timelike vectors tangent to the
horizon, but a horizon is a null surface, and timelike vectors cannot
be tangent to null surfaces, so $r=r_H$ cannot be a true event
horizon\footnote{We thank C.~A.~R.~Herdeiro for discussions on this
point.}.)  We leave a more detailed analysis of the general solution
for the future.

Using our base manifold (\ref{linebundle}), constructing
generalizations of the black holes of \cite{Gutowski:2004ez} to the
case of flat ($k=0$) or hyperbolic ($k=-1$) horizons should be
straightforward.  It would also be interesting to obtain the most
general function $V(r)$ in the line bundle (\ref{linebundle}) that has
$\Theta = 0$.  Unfortunately this condition leads to a rather
complicated fifth order differential equation for $V(r)$ that we were
not able to solve in general. Another open question is whether the
rotating black holes of \cite{Klemm:2000vn} (which also have
CTCs \cite{Caldarelli:2001iq}) can be described by using
the base (\ref{linebundle}), and if so, whether they still have
$\Theta = 0$.

\section{The G\"odel-de~Sitter universe}

\label{GdS}

We close this paper by presenting a generalization of the G\"odel-type
universe of \cite{Gauntlett:2002nw} to the case $\Lambda < 0$ (which,
with our signature, corresponds to de~Sitter). As it has been observed
in \cite{Kastor:1992nn,London:ib}, one can embed asymptotically flat
supersymmetric black holes into a de~Sitter space by introducing a
specific time dependence, which is either a time exponential
multiplying the radial coordinate or simply an appropriate linear
function in time added to the harmonic function.  This procedure,
which breaks of course supersymmetry, can also be done for the case at
hand yielding for the metric and gauge field (that solve the equations
of motion),
\begin{eqnarray}
ds^2 &=& f^2(dt + \omega)^2 - f^{-1}(dr^2 + \frac{r^2}4[(\sigma_1^L)^2
         + (\sigma_2^L)^2 +(\sigma_3^L)^2])\,, \nonumber \\ 
A &=& \frac{\sqrt 3}2f[dt - {\cal F}_1 r^2 \sigma_1^L + {\cal F}_2
      r^2 \sigma_2^L]\,, \nonumber \\ \omega &=& - {\cal F}_1 r^2
      \sigma_1^L + {\cal F}_2 r^2 \sigma_2^L\,, \nonumber \\ 
f^{-1} &=& \sqrt{\frac{-2\Lambda}3} \, t + {\cal H}\,, \label{goedeldS}
\end{eqnarray}
where ${\cal H}$ denotes an arbitrary harmonic function on the {\em
flat} base manifold. This metric has a curvature singularity at
$f^{-1} = 0$, which gives an $r$-dependent lower bound for the time
$t$. This repulson-type initial singularity is the starting point of
the eternal expanding multi black hole space-time. There is also a
time reversal, collapsing solution, which has a future singularity and
can be obtained by changing the sign in front of $t$ in the function
$f$.  For $\Lambda = 0$, it reduces to a solution found in
\cite{Gauntlett:2002nw} with the G\"odel universe corresponding to
${\cal H}=1$. Unlike the AdS case, for $t\to\infty$ the geometry
(\ref{goedeldS}) does approach de~Sitter space. The G\"odel
deformation ($\sim {\cal F}_{1/2}$) yields again regions with CTCs,
which are now time dependent. In fact, on slices with $dt=dr=0$, CTCs
occur whenever
\begin{equation}
\label{233}
4 r^2({\cal F}_1^2 + {\cal F}_2^2) > f^{-3}\,,
\end{equation}
and since $f$ is a time- and radial-dependent function, this relation
defines a shell which moves through spacetime with increasing velocity
$(\dot r)_0$ towards larger values of $r$. In fact, as $f$ goes to
zero for $t\to\infty$, this means that on any slice given by $dt=0$
the CTCs disappear for sufficiently large $t$. On the other hand, if
we go back in time, any point of space enters the region with CTCs.
Another question is whether for a given time any point is part of a
CTC. This would be the case if one can travel into the region of
CTCs, go back in time there and return to the starting point. As long
as we are outside the region defined by (\ref{233}), $t$ is a
good time coordinate. Let us consider radial velocities
only. From the metric one finds as maximal radial velocity $(\dot
r)_{max} = f^{3/2}$, which has to be compared with the velocity of
the region of CTCs as given in (\ref{233}). For ${\cal H} = 1 +
\frac{\mu}{r^2}$ we find, that whenever the relation
\begin{equation}
(\dot r)_{0}= 
\frac{\sqrt{-6 \Lambda}}{ 4 \sqrt{ ({\cal F}_1^2 + {\cal F}_2^2) \, f\, }
  + 6 \frac{\mu }{ r^3}} > f^{3/2}
\end{equation}
holds, a radial signal cannot reach the region of CTCs and 
therefore, regions satisfying this relation are free of CTCs.


\section*{Acknowledgements}
\small
K.~B.~would like to thank the Theory group of the University of
Milan, where part of this work has been carried out. This work was
partially supported by INFN, MURST and by the European Commission RTN
program HPRN-CT-2000-00131. The work of K.~B.~is supported by a DFG
Heisenberg grant. We are grateful to C.~A.~R.~Herdeiro for
useful comments.
\normalsize


\end{document}